\documentclass[twocolumn,superscriptaddress,prl,aps]{revtex4}
\usepackage{amssymb}

\usepackage{amsfonts}
\usepackage{amsmath}
\usepackage{epsfig}

\begin{document}

\title{Deep Learning for Feynman's Path Integral in Strong-Field Time-Dependent Dynamics}

\author{Xiwang Liu}

\affiliation{Research Center for Advanced Optics and Photoelectronics, Department of Physics, College of Science, Shantou
University, Shantou, Guangdong 515063, China}
\affiliation{Department of Mathematics, College of Science, Shantou
University, Shantou, Guangdong 515063, China}

\author{Guojun Zhang}

\affiliation{Research Center for Advanced Optics and Photoelectronics, Department of Physics, College of Science, Shantou
University, Shantou, Guangdong 515063, China}

\author{Jie Li}

\affiliation{Research Center for Advanced Optics and Photoelectronics, Department of Physics, College of Science, Shantou
University, Shantou, Guangdong 515063, China}

\author{Guangluo Shi}

\affiliation{Research Center for Advanced Optics and Photoelectronics, Department of Physics, College of Science, Shantou
University, Shantou, Guangdong 515063, China}

\author{Mingyang Zhou}
\affiliation{Research Center for Advanced Optics and Photoelectronics, Department of Physics, College of Science, Shantou
University, Shantou, Guangdong 515063, China}

\author{Boqiang Huang}
\affiliation{Mathematisches Institut, Universit\"{a}t zu K\"{o}ln, 50931, K\"{o}ln, Germany}

\author{Yajuan Tang}
\affiliation{Department of Electronic and Information Engineering, College of Engineering, Shantou
University, Shantou, Guangdong 515063, China}

\author{Xiaohong Song}
\affiliation{Research Center for Advanced Optics and Photoelectronics, Department of Physics, College of Science, Shantou
University, Shantou, Guangdong 515063, China}
\affiliation{Department of Mathematics, College of Science, Shantou
University, Shantou, Guangdong 515063, China}
\affiliation{Key Laboratory of Intelligent Manufacturing Technology of MOE, Shantou University, Shantou, Guangdong 515063, China}

\author{Weifeng Yang}
\email{wfyang@stu.edu.cn}
\affiliation{Research Center for Advanced Optics and Photoelectronics, Department of Physics, College of Science, Shantou
University, Shantou, Guangdong 515063, China}
\affiliation{Department of Mathematics, College of Science, Shantou
University, Shantou, Guangdong 515063, China}
\affiliation{Key Laboratory of Intelligent Manufacturing Technology of MOE, Shantou University, Shantou, Guangdong 515063, China}

\date{\today}

\begin{abstract}
Feynman's path integral approach is to sum over all possible spatio-temporal paths to reproduce the quantum wave function and the corresponding time evolution, which has enormous potential to reveal quantum processes in classical view. However, the complete characterization of quantum wave function with infinite paths is a formidable challenge, which greatly limits the application potential, especially in the strong-field physics and attosecond science. Instead of brute-force tracking every path one by one, here we propose deep-learning-performed strong-field Feynman's formulation with pre-classification scheme which can predict directly the final results only with data of initial conditions, so as to attack unsurmountable tasks by existing strong-field methods and explore new physics. Our results build up a bridge between deep learning and strong-field physics through the Feynman's path integral, which would boost applications of deep learning to study the ultrafast time-dependent dynamics in strong-field physics and attosecond science, and shed a new light on the quantum-classical correspondence.

\end{abstract}

\maketitle

The wave function and the temporal evolution contain all information of quantum physics. However, they might be possibly the hardest to grasp in the classical world. Seventy years ago, Feynman proposed a path integral approach which has been viewed as the ``sum over paths or histories" version of quantum mechanics, i.e. the wave function can be represented as a coherent superposition of contributions of all possible spatio-temporal paths \cite{Feynman1948RMP,Dyson1980}. Even though the Feynman's path integral (FPI) has been considered as the most fundamental way to interpret the quantum mechanics and answer what is the nature of measurements, the complete characterization of quantum wavepacket with all possible paths is formidable due to track ergodicity. Typically, only a very limited amount of paths could be accessed, and therefore only a reduced amount of information of quantum wavepacket could be obtained in different approximation methods so far.

The development history of semiclassical methods based on FPI in strong-field physics, from the strong-field approximation (SFA) to the Coulomb corrected strong-field approximation (CCSFA) and quantum trajectory Monte Carlo methods, also proves that the more trajectories have been adopted, the more information could be extracted \cite{Keldysh1965,Faisal1973,Reiss1980,Lewenstein1994,Salieres2001Science,Kopold2000,Schafer2004,Blaga2009NatPhys,Faisal2009NatPhys,Quan2009,Huismans2011,Walt2017NC,Yan2010PRL,Xiao2016,Li2014PRL,Shilovski2016,Geng2015PRL,LiuMM2016PRL,Song2017,Song2016,Yang2016,Faria2019}. As a result, despite of the notable success of these methods, there still exist a large number of unexplored regimes, including the open question about whether one could truly achieve the quantum-classical correspondence. Actually,
with increasingly sophisticated experiments, the limitation of existing semiclassical methods based on FPI for reproducing and explaining some quantum phenomena has been becoming increasingly evident due to the limited amount of paths, especially for the new attosecond measurements where a series of high-resolution photoelectron spectra with different pump-probe delays are needed to obtain attosecond time-resolved movies of electrons \cite{Porat2018,Gong2017, Song2018,Zipp2014,Leone2014,Razourek2015,Li2019,Liu2019}.

Since the game Go was mastered by deep neural networks (DNNs), deep learning (DL) has received extensive attention \cite{Silver2016,Silver2017}. Recently, this technique has powered many fields of science, including planning chemical syntheses \cite{Segler2018}, acceleration of super-resolution localization microscopy and nudged elastic band calculations \cite{Wei2017,Nehme2018,Strack2018,Torres2019}, classifying scientific data \cite{Webb2018,Gabbard2018}, solving high-dimensional problems in condensed matter systems \cite{Carleo2017,Carrasquilla2017,Nieuwenburg2017,Torlai2018,Xia2018,Zhang2018,Sarma2019}, reconstructing the shape of ultrashort pulses \cite{Zahavy2018}, and so on. However, to our knowledge, its power in strong-field physics has not yet been excavated. As a result, it is very important to figure out: (i) whether and how DL could be used to solve the problems in strong-field physics and attosecond science? (ii) Could DL help discover new physics in these rapidly developed fields?

Here we demonstrate that FPI provides a good breakthrough point for DL to inroad strong-field physics and attosecond science. We identify that, assisted by a multilayer perceptron (MLP) to pre-classify data of electron trajectories \cite{Goodfellow2016BOOK}, the proposed DNNs can be trained with only input and output data of few available sample trajectories without knowing in advance the detailed dynamics of the sample trajectories, and then create a predictive tool which directly predict the final results of arbitrary trajectory given its initial conditions. This allows to perform the FPI in strong-field physics using DL-predicted data, which would break the bottleneck of conventional semiclassical methods due to large computation cost of ergodic tracing trajectories.
The high-efficiency of deep-learning-performed strong-field Feynman's formulation (DLPSFFF) thus allows us to tackle really tough tasks in which huge amounts of data are far beyond processing capacity of existing methods in strong-field physics. Moreover, we show that the MLP itself is a powerful tool to classify trajectories and helps reveal the underlying physics. Our results provide a promising technique not only for exploring the new physics of ultrafast electron dynamics in attosecond measurements, but also for approaching the limit of the quantum-classical correspondence in the fundamentalism of FPI.

\begin{figure}
\includegraphics[width=0.5\textwidth]{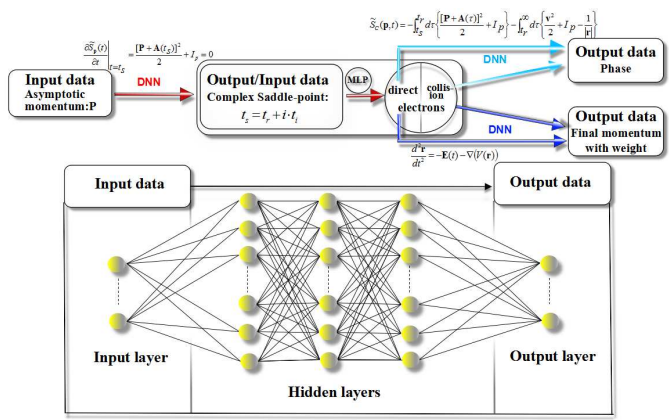}
\caption{\textbf{DNNs encoding the time-dependent dynamics in photoemission.} Schematic illustrations of DL to learn the
mapping relationships of saddle-point of quantum complex emission time, finial momentum and phase, respectively. Fully-connected multi-layer neural networks perform both of the forward propagation and
 back-propagation algorithms which repeatedly adjust the weights of the connections to minimize the loss.}

\label{Figure1}
\end{figure}

We take CCSFA as an example since it has shown broad prospect in interpretation of strong-field experiments \cite{Huismans2011,Porat2018,Yan2010PRL,Geng2015PRL}. It should be noted that the proposed strategy is not limited to CCSFA, but can be extended to other approaches based on FPI in strong-field physics and attosecond science. The architecture of the DNNs is shown in Fig. \ref{Figure1}. The first DNN provides initial conditions of electrons emerging in the continuum after tunneling. The second and third ones predict the phase and final momentum, respectively.

In the CCSFA method, the initial conditions of electron trajectories appearing in the continuum after quantum tunnelling are obtained by solving the saddle-point equation (see Supplemental Material \cite{SM}). In this stage, the input data are a series of asymptotic momentum $\mathbf{P}$, and the output data are the real part and imaginary part of the saddle-point $t_s=t_r+i\cdot$$t_i$, respectively. Usually, it is very time consuming to search the saddle-points in a complex time plane, particularly for complicated laser electric fields. Whereas, using DL, this can be easily done with a high accuracy and an ignorable cost (see Supplemental Material \cite{SM}). For other semiclassical methods, like time-sampling CCSFA and QTMC methods where the initial conditions of space-time paths including its weight can be obtained by other methods \cite{Xiao2016,Li2014PRL,Shilovski2016,Song2016}, this step can be skipped.

The main challenge for applying DL to perform strong-field FPI lies in whether DNNs could blindly learn and predict the complex ultrafast time-dependent electron dynamics from a limited of raw sample data without knowing any mathematical functions in advance. To this end, both the final momentum and phase of path should be accurately predicted. In our work, the training set consists of $5\times10^{5}$ sample trajectories which can be quite easily obtained by numerical solution of time-dependent Newton equation (TDNE) with Coulomb interaction and the path integral in CCSFA (see Supplemental Material \cite{SM}). The test set consists of $1\times10^{5}$ trajectories which are not included in the training set. It is much more difficult for DNNs to learn the mapping relationship of final phase since it is the integral over each time step along a temporal and spatial path. In the following, we mainly show the results about the phase $S(\textbf{p},t_{s})$.

We find that, for simple cases, e.g. SFA in which Coulomb interaction is neglected, a simple artificial neural network (ANN) with only one hidden layer with 30 neurons can be well trained to reproduce the test results (see Supplemental Material \cite{SM}). This consists with the conclusions of very recently work that for simple systems like a pendulum, the ANN could recover similar representation with one-dimensional TDNE only from given samples \cite{Iten2020}. Moreover, it has been demonstrated that ANNs can be used to solve differential equations for applications to the calculation of cosmological phase transitions \cite{Piscopo2019}. However, here we find that when the Coulomb potential and the spatial gradient are fully accounted contains and the TDNE contains the derivatives with respect to both time and space, DNNs fail to learn the map relationships even though 3 hidden layers with 250 neurons in each layer are adopted. Most of predicted values deviate from the true ones obtained by CCSFA simulation (see Fig. 2(g)). Further increasing neurons and hidden layers could not improve the performance of DL.

To find out the reasons for such failure due to the complexity of Coulomb interaction, we analyze the final phase distributions with initial momenta in the training and test data sets. It can be found that the phase distribution is very flat and smooth over a large area [denoted by the arrow ``1" in Fig. 2(a)], which is very similar with that in SFA result (see Supplemental Material \cite{SM}). The sample analysis indicates that these are direct electron trajectories which do not revisit the parent ion after emerging in the continuum [see the left inset of Fig. 2(h)]. However, some peaks and obvious fluctuation can be seen in the area [denoted by the arrow ``2" in
Fig. 2(a)], which is absent in SFA result. The sample analysis shows that these electrons undergo at least one collision with the ion as shown in the left inset of Fig. 2(i). It is physically reasonable that during the process of collision, the electrons are very close to the Coulomb singularity, and a tiny difference in spatial position would have a huge impact on the final momentum and phase due to the large spatial gradient near the singularity. All these indict that the data distribution of initial states have indeed already encode the information of time-dependent dynamics. Moreover, the features of data distributions are quite different for different type of trajectories. Obviously, the data of direct and rescattered electron trajectories with different dynamic processes are mixed up in the set. Therefore, we infer that due to the diversity of data, DNN cannot find a unified mapping relationship between initial and final states for different type of space-time paths.

To overcome this challenge, we propose to employ multiple DNNs to learn different mapping relationships and predict directly the final results for different type of trajectories. However, to perform this DL strategy, one should be up against another difficulty that is to divide the data of initial conditions of arbitrary electron trajectory into different groups without knowing the time-dependent dynamics in advance. For conventional calculation strategy, it is impossible that classifying trajectories only with the data of initial conditions without tracing the space-time paths. Here we construct a full connected multilayer feed forward network, known as MLP, with three hidden layers to classify the data of direct and scattered trajectories. Similar network has recently been used to identify phase of condensed-matter \cite{Carrasquilla2017}.

To train the MLP, the sample data need to be labelled firstly. We label ``$1$" for the direct trajectories and ``$-1$" for the scattered ones during the course of sample simulation. The training process is to build a mapping relationship between the input, i.e. the initial conditions of the trajectories, and their labels. After being trained, the MLP will output a value between $-1$ to $1$ for given initial conditions of arbitrary trajectory. Figure 2(e) shows the output label values as a function of the input initial momentum. It can be seen that most of the output values have already been $-1$ or $1$, and some discrete points distribute at the margins of these two parts. By examining the trajectories corresponding to these discrete points, we find that these trajectories are slightly influenced by the Coulomb potential and indeed intermediate between scattered and direct trajectories. Here we classify the test data completely by choosing a threshold 0 and setting to ``$1$" when the output label value is larger than or equal to 0, otherwise, setting to ``$-1$". Comparing with the true label distribution of the test set [see Fig. 2(d)], the MLP works very well [see Fig. 2(f)], which demonstrates that the characters of trajectories with different dynamics have already been encoded in the data of initial conditions, and the MLP can extract the information of electron dynamics and identify different type of trajectories only from the initial conditions [see Figs. 2(b) and (c)].

After the pre-classification, we then feed the classified input data into the corresponding DNNs trained with the two subsets of direct and scattered samples, respectively, and compare the predicted outputs with the true values in the test subset. One can see that both of the two DNNs learn very well [see Figs. 2(h) and 2(i)], and the errors, i.e. the difference between the true and predicted phase $S(\textbf{p},t_{s})_{true}-S(\textbf{p},t_{s})_{DL}$, for both direct [the right inset of Fig. 2(h)] and rescattered [the right inset of Fig. 2(i)] electron trajectories are located mainly around 0. Same procedure can be applied to train DNNs for learning the final momentum of photoelectrons.

\begin{figure}
\includegraphics[width=1.0\columnwidth]{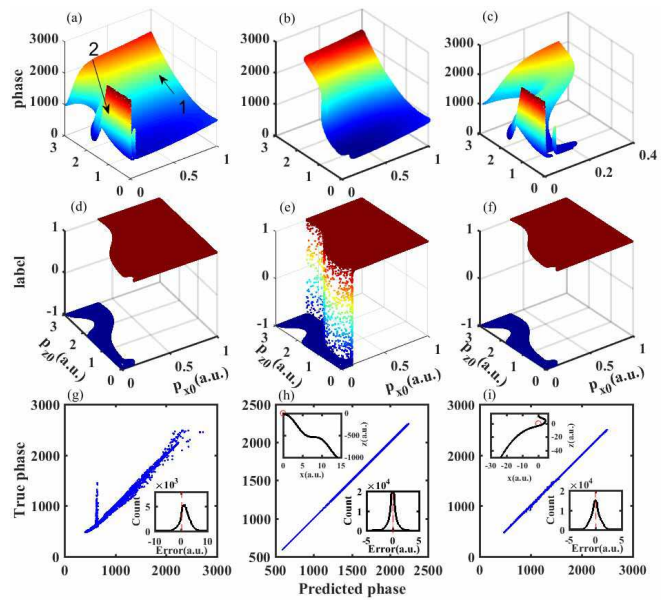}
\caption{\textbf{Deep learning of phase.} (a) The test data set, (b) and (c) two labeled subsets after classification with MLP (see the text). The black numbers and arrows in (a) denote different mapping relationships in the data set. (d) the true label distribution of the test set, (e) the original output label values by MLP, (f) the final classification of the test set by MLP. (g), (h) and (i) Histograms of the true vs predicted phases for test data corresponding to (a), (b) and (c). The right insets in (g), (h) and (i) show the distribution of error away from the perfect predictions. The left insets in (h) and (i) show the typical electron trajectories in the corresponding data. The parameters are: a argon atom with $I_p=-0.579$ a.u., was ionized by a linearly polarized ultrashort few-cycle laser pulse at wavelength of 800 nm and peak intensity of $1\times10^{14}$$\mathrm{W/cm}^{2}$.}

\label{Figure2}
\end{figure}

To test the validity of the DLPSFFF, we adopt the ultrafast ionization of a hydrogen atom subject to an elliptically polarized laser pulse with a ellipticity 0.88 at wavelength of 800 nm and peak intensity of $1\times10^{14}$$\mathrm{W/cm}^{2}$  [see Figs. 3(a)-3(e)] as an example, which is the same parameters of recent experiment \cite{Sainadh2019} to determine `tunnelling times', the fundamental issue of quantum mechanics and attosecond science.

Figure 3(a) shows the photoelectron momentum distribution (PMD) constructed with only the sampled training data of $5\times10^5$ trajectories. Actually, these samples are very few, which are even not enough to form interference structures. After being trained only with data of these few sample trajectories, the DNNs can predict directly the final momentum and phase of arbitrary trajectory with negligible calculation cost. Figure 3(c) shows the PMD constructed with the DL-predicted data. For comparison, the number of trajectories in the DLPSFFF is same with that in the CCSFA simulation [Fig. 3(b)]. Moreover, we also present the quantum result, i.e. the numerical solution of time-dependent Sch\"{o}rdinger equation (TDSE), as benchmark [see Fig. 3(d)]. The DL-predicted PMD [Fig. 3(c)] agrees well with the CCSFA and TDSE results [see Figs. 3(b) and 3(d)] all of which show the above-threshold ionization peaks, one of typical interference structures in strong-field photoionization \cite{Agostini1979,Becker2002}. Especially, the DL-predicted photoelectron angular distribution reproduces exactly the CCSFA result and the numerical solution of TDSE [Fig. 3(e)], which demonstrated the excellent performance of the DLPSFFF.

\begin{figure*}
\includegraphics[width=1.2\columnwidth]{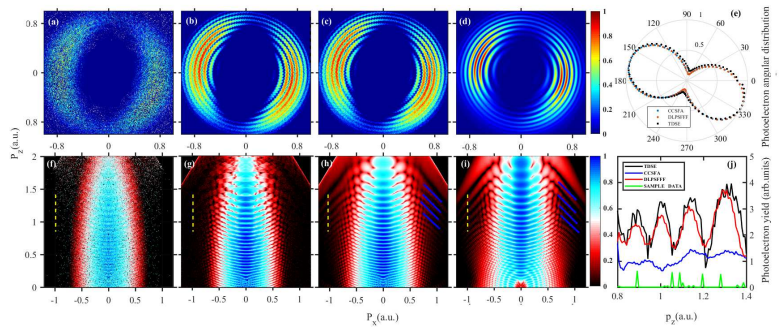}
\caption{\textbf{Comparison between the conventional simulations and the DLPSFFF predictions.} The PMDs constructed with (a)(f) $5\times10^{5}$
sample training data (b)(g) $1\times10^{8}$ trajectories simulated by original CCSFA treatment, (c) $1\times10^{8}$ and (h) $1\times10^{10}$ data predicted by DLPSFFF, and (d)(i) quantum TDSE results. (e) The comparison of photoelectron angular distributions simulated by CCSFA and TDSE with that predicted by DLPSFFF. (j)  the photoelectron yield along $\mathbf{p_{x0}}$=-1 a.u. (the yellow dash line) in (f), (g), (h) and (i). The parameters used in simulations are: upper panel, a hydrogen atom was ionized by a elliptically polarized laser pulse with a ellipticity 0.88 at wavelength of 800 nm and peak intensity of $1\times10^{14}$$\mathrm{W/cm}^{2}$ lower panel, the same as in Fig. 2.}

\label{Figure3}
\end{figure*}

It should be stressed that the DL-prediction strategy is much faster than ergodic simulation in conventional methods. For example, tracing $1\times10^{4}$ trajectories one by one in CCSFA simulation spends 204.4 seconds on a single CPU processor, while the prediction of same number of data by DLPSFFF only takes 0.43 seconds. Actually, the more trajectories are needed, the more powerful performance of DLPSFFF shows.

The high-performance of the DLPSFFF provides unprecedented opportunities to attack unsurmountable tasks by usual semiclassical methods and help uncover new physics. Figures 3(f)-3(g) show another example. The quantum TDSE simulation shows clearly an undetected oblique interference structure [denoted by blue solid lines in Fig. 3(i)] which is absent in the usual semiclassical CCSFA simulation even with $10^{8}$ trajectories [see Fig. 3(g)]. It should be noted that the calculation cost of simulating $10^{8}$ trajectories in semiclassical methods has already been very large \cite{Porat2018}, but these are actually a very small fraction of the all possible paths required by the fundamentalism of FPI. As a result, it is quite reasonable that some new physics might be lost due to reduced amount of trajectories in conventional simulations.

To reproduce the TDSE result, we trained the DNNs with $5\times10^{5}$ sample data of trajectories. Again, the quantum interference was hardly recognised in the constructed PDM [see Fig. 3(f)]. After trained with these few available sample data, it is quite easy for the DLPSFFF to predict directly the final momentum and phase for $1\times10^{10}$ trajectories [see Fig. 3(h)]. The predicted PMD now clearly shows the oblique interference structure in TDSE result. If ergodic tracing all these trajectories with conventional semiclassical methods, the calculation cost would be huge due to the exponential increase of the number of paths. Figure 3(j) shows the photoelectron yields along the yellow dash line for different cases in Figs. 3(f)-3(i). For the case only with sample data of classical trajectories (green line) and the semiclassical simulation with $10^{8}$ trajectories (blue line), only noise can be detected. Whereas, good agreement is achieved between the DLPSFFF result constructed with $1\times10^{10}$ predicted data of trajectories (red line) and the TDSE result (black line). As a result, the DLPSFFF provides a powerful tool to push toward the limit of the classical-quantum correspondence and the fundamentalism of FPI.

Moreover, we show that the MLP and pre-classification scheme can help reveal the underlying physics. We demonstrated that the oblique interference structure originates from the interference of large-angle forward-scattering electron trajectories (see Supplemental Material \cite{SM}). Specifically, the interference of electrons with different scattering angles but same final momentum induces this novel structure. These large-angle forward-scattering trajectories are very closed to the core during the process of scattering, hence their proportion in the total trajectories is relatively small. Only when the number of the total trajectories is very huge, the interference between them could be visible. Since these trajectories are very close to the core, they could carry the information about parent ion. Therefore, the predicted interference fringes and thus the underlying physics might be used to image atomic and molecular spatio-temporal dynamics which needs to be investigated in the future work.

In summary, we introduce a computational strategy that utilizes DL to implement Feynman's formulation in strong-field physics, therefore getting over the drawback of inherent brute-force calculation of existing methods. Our results demonstrate that DNNs can be well trained with a very small number of samples which can only constitute a fuzzy outline of observable. Once being trained, the DLPSFFF can predict directly the final results for as many as trajectories required in reconstructing high-resolution spectra. Moreover, our results show that the pre-classification scheme is an efficient way to tackle the complicated problems where trajectories are very diverse, and uncover the underlying physics in strong-field physics. The feasibility study in this work would unlock the great potential of combined DL and FPI in analyzing and predicting strong-field experimental phenomena, which will lead to not only overcoming challenges beyond the today's computational capacity and methods, but also touches on the fundamental issue of the quantum-classical correspondence.

\section{Acknowledgment}
We thank H Huang, S Ran, H. Zhai, J. Yao, and J. Liu for the fruitful discussions. The work was supported by the National Natural Science Foundation
of China (Grant No. 91950101, No. 11674209, No. 11774215 and No. 11947243), Sino-German Mobility Programme (Grant No. M-0031), Department of Education of Guangdong Province (Grant No.2018KCXTD011), High Level University Projects of Guangdong Province (Mathematics, Shantou University), and the Open Fund
of the State Key Laboratory of High Field Laser Physics (SIOM).

\end{document}